\documentclass[aip,
 amsmath,amssymb,
 graphicx,reprint,]{revtex4-2}

\usepackage{graphicx}% Include figure files
\usepackage{dcolumn}% Align table columns on decimal point
\usepackage{bm}% bold math
\usepackage{xcolor, color, soul}
\usepackage{MnSymbol}

\begin{document}

\title{Temperature Dependence of Magnetic Anisotropy and Domain Wall Tuning in BaTiO$_3$(111)/CoFeB Multiferroics}% Force line breaks with \\

\author{R. G. Hunt}
\affiliation{School of Physics and Astronomy,
            University of Leeds, LS2 9JT, United Kingdom}
\affiliation{Bragg Centre for Materials Research, University of Leeds, LS2 9JT, United Kingdom}

\author{K. J. A.  Franke}
\affiliation{School of Physics and Astronomy,
            University of Leeds, LS2 9JT, United Kingdom}
            
\author{P. S. Keatley}
\affiliation{Department of Physics and Astronomy,
            University of Exeter, Stocker Road, Exeter, EX4 4QL, United Kingdom}
            
\author{P. M. Shepley}
\affiliation{School of Physics and Astronomy,
            University of Leeds, LS2 9JT, United Kingdom}
\affiliation{Bragg Centre for Materials Research, University of Leeds, LS2 9JT, United Kingdom}
            
\author{M. Rogers}
\affiliation{School of Physics and Astronomy,
            University of Leeds, LS2 9JT, United Kingdom}
\affiliation{Bragg Centre for Materials Research, University of Leeds, LS2 9JT, United Kingdom}

\author{T. A. Moore}
\affiliation{School of Physics and Astronomy,
            University of Leeds, LS2 9JT, United Kingdom}  
\affiliation{Bragg Centre for Materials Research, University of Leeds, LS2 9JT, United Kingdom}

\begin{abstract}

Artificial multiferroics consist of two types of ferroic materials, typically a ferroelectric and ferromagnet, often coupled interfacially by magnetostriction induced by the lattice elongations in the ferroelectric. In BaTiO$_3$ the magnitude of strain induced by these elongations is heavily temperature dependent, varying greatly between each of the polar crystal phases and exerting a huge influence over the properties of a coupled magnetic film. Here we demonstrate that temperature, and thus strain, is an effective means of controlling the magnetic anisotropy in BaTiO$_3$(111)/CoFeB heterostructures. We investigate the three polar phases of BaTiO$_3$: tetragonal (T) at room temperature, orthorhombic (O) below 280 K and rhombohedral (R) below 190 K, across a total range of 77 K to 420 K. We find two distinct responses; a step-like change in the anisotropy across the low-temperature phase transitions, and a sharp high-temperature reduction around the ferroelectric Curie temperature, measured from hard axis hysteresis loops. Using our measurements of this anisotropy strength we are then able to show by micromagnetic simulation the behaviour of all possible magnetic domain wall states and determine their scaling as a function of temperature. The most significant changes occur in the head-to-head domain wall states, with a maximum change of 210 nm predicted across the entire range effectively doubling the size of the domain wall as compared to room temperature. Notably, similar changes are seen for both high and low temperatures which suggest different routes for potential control of magnetic anisotropy and elastically pinned magnetic domain walls.

\end{abstract}

\maketitle

%\tableofcontents

\section{Introduction}

Ferroelectric and multiferroic materials are becoming an increasingly important set of materials for next-generation devices due to the ability to manipulate non-volatile states by application of electric fields\cite{Bea2008, Ajayan2023}. The integration of these materials into spintronic devices offers the potential to improve existing device designs by replacing current-controlled operations with electric field-controlled ones which could reduce the energy required to operate devices by orders of magnitude\cite{Franke2015, Yin2017}. 

Associated with the ferroelectric order is a substantial lattice elongation, on the order of 1\% of the in-plane lattice parameter, that makes these materials suitable for imparting a large strain. Conventional magnetic multilayer devices, coupled to a ferroelectric material, can be modified and controlled by applying an electric field to modify the strain, lowering the power needed to operate a device\cite{Damjanovic1998}. This makes ferroelectrics a core material in the field of straintronics\cite{Roy2011, Bukharaev2018} in which devices can be modified or manipulated by strain. The scope of these modifications is very broad, but the applications in which ferroelectrics play a particularly large role include thin film systems where ferroelectric materials can be used to control the interfacial strain. Indeed, research involving piezoelectric and ferroelectric materials has already shown that it is possible to control magnetic domain wall velocities\cite{Shuai2022}, domain structure\cite{Lahtinen2012}, and the magnitude of micromagnetic parameters such as effective anisotropy\cite{Shepley2015}.
 
\begin{figure}
    \centering
    \includegraphics[width=0.8\columnwidth]{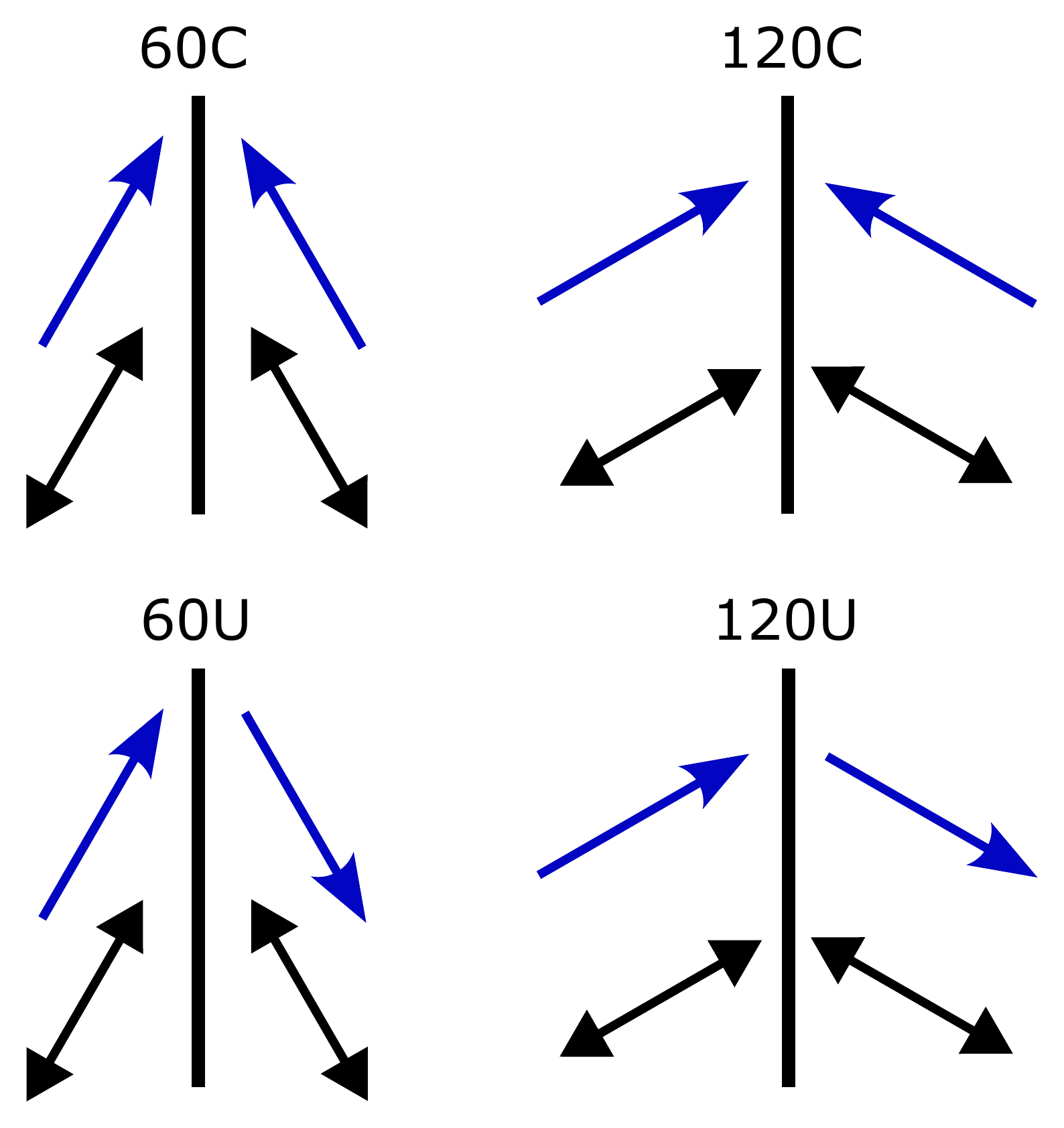}
    \caption{Schematic illustration of the magnetization and anisotropy configurations in the strain-coupled ferromagnet. The domain wall configurations are labelled by the underlying angle between the magnetoelastic anisotropy axes (60$^\circ$ or 120$^\circ$) and the charged or uncharged nature of the domain wall structure leading to four configurations: 60U, 60C, 120U and 120C. Black double-headed arrows indicate the direction of magnetoelastic anisotropy, blue arrows represent the direction of local magnetization.}
    \label{dw_configs}
\end{figure}

In our previous work\cite{Hunt2023}, we have shown that growing a ferromagnetic film of Co$_{40}$Fe$_{40}$B$_{20}$ on a (111)-oriented BaTiO$_3$ substrate couples the magnetic domain structure to the ferroelectric domain structure and leads to two in-plane magnetoelastic anisotropy configurations (illustrated in Fig. \ref{dw_configs}) in which the imprinted magnetoelastic anisotropy rotates by either 60$^{\circ}$ or 120$^{\circ}$ between adjacent ferroelectric domains, and that this impacts the magnetic domain structure and the magnetic field response of magnetic domain walls, giving rise to the 60U, 60C, 120U and 120C domain wall states. Previous investigations of this system were limited to static effects at room temperature and without applied electric field. In this paper we will detail the effects of temperature.

BaTiO$_3$ (BTO) exhibits three crystal phase transitions from rhombohedral to orthorhombic at 190 K, orthorhombic to tetragonal at 280 K, and tetragonal to cubic at 400 K\cite{Acosta2017} (dependent on crystal quality and if the sample is being heated or cooled), with the rhombohedral (\textbf{R}), orthorhombic (\textbf{O}), and tetragonal (\textbf{T}) crystal phases all displaying ferroelectric order. Each of these phases has different preferred orientations of ferroelectric polarization and so these phase transitions are accompanied by large changes in the lattice parameters of BTO, particularly along the directions of ferroelectric polarization where the lattice is elongated. Interestingly, the projection of polarization onto the (111) plane in all phases results in the same two possible rotations of in-plane projection of the polarization, where the polarization rotates by either 60$^\circ$ or 120$^\circ$ through the domain wall. This angle directly affects the magnetic domain wall angle, reducing the rotation of magnetization between adjacent domains and creating the possibility for the domain walls to be either head-to-head (charged) or head-to-tail (uncharged). 

Here, we investigate the temperature response of coupled BTO(111)/CoFeB films both in bulk and at the individual domain level. By sweeping the temperature we can access different polar phases with large changes in lattice elongation and assess the viability of strain-dependent devices based on BTO(111) substrates, measured by the response of the properties of the ferromagnetic film deposited on top. Similar to the work performed in the literature\cite{Lahtinen2013}, we find that the magnetic thin film remains coupled in all crystal phases regardless of the change in strain.We also determine values for the magnetoelastic anisotropy in CoFeB within individual coupled ferroelectric domains across the entire temperature-dependent structural range for BTO, from 77 K (below which we expect no significant changes in the lattice constants) to 417 K (above which there is no ferroelectric order). Finally, we perform micromagnetic simulations of the domain wall width to show the impact these changes represent and find that at the very extremes of the temperature range the difference between domain wall configurations becomes significant for charged domain wall structures. This will have important implications for devices making use of them such as domain wall resonators\cite{Hamalainen2018, VandeWiele2016}.  

\section{Methods}

Samples were grown by DC sputtering onto commercially available BaTiO$_3$(111) substrates purchased from SurfaceNet. These substrates display ferroelectric domains as received and preparation involved cleaning by sonication in acetone and then isopropanol. Using the Royce deposition system\cite{Royce} at Leeds, a ferromagnetic Co$_{40}$Fe$_{40}$B$_{20}$ layer is deposited at 300$^\circ$C above the Curie temperature of the substrate to promote more efficient strain transfer. The full sample structure is BaTiO$_3$(111)/Co$_{40}$Fe$_{40}$B$_{20}$(20nm) @ 300$^\circ$C/Pt(5nm) @ 27$^\circ$C. As in our previous work, the as-received domain structure of the BTO is diverse and contains regions in which the polarization rotates through either 60$^\circ$ or 120$^\circ$ between adjacent ferroelectric domains. This results in a diversity of the accompanying magnetic configurations once the ferromagnetic layer is deposited. More details can be found in our previous work\cite{Hunt2023}.

Magnetometry was performed using a Quantum Design MPMS3 superconducting quantum interference device-vibrating sample magnetometry (SQUID-VSM) to investigate the bulk changes in magnetic moment and magnetization over a temperature range of 77-300 K in magnetization v temperature (MvT) measurements. Values of saturation magnetization were extracted from $\pm$ 1 T magnetic hysteresis loops performed at the respective temperatures with background contributions subtracted. A temperature-dependent Bloch $T^{\frac{3}{2}}$ fit\cite{Argyle1963} to the data is performed from the values of $M_{sat}$ obtained and used to extrapolate values of $M_{sat}$ between these data points and above room temperature up to 417K. MvT measurements are presented with no background contributions removed. 

Local hysteresis measurements below room temperature were carried out using a wide-field Kerr microscope\cite{Soldatov2017} with an optical cryostat at the EXTREMAG\cite{Extremag} facility in Exeter with a 60$\times$ objective lens. The sample is cooled to a base temperature of 77 K and then warmed to achieve the desired temperature. Magnetic hysteresis loops at each temperature are taken along the magnetic hard axis of an individual region coupled to a ferroelectric domain and the anisotropy field is calculated from the resulting loop. Measurements above room temperature were performed at Leeds with a heater stage in the same manner. All measurements are taken in the positive (warming) temperature direction.

\section{Results}

%% Cut fig 2 and 3, show instead that the transition is suppressed with higher fields - 1T sweep example.

\begin{figure}
    \centering
    \includegraphics[width=\columnwidth]{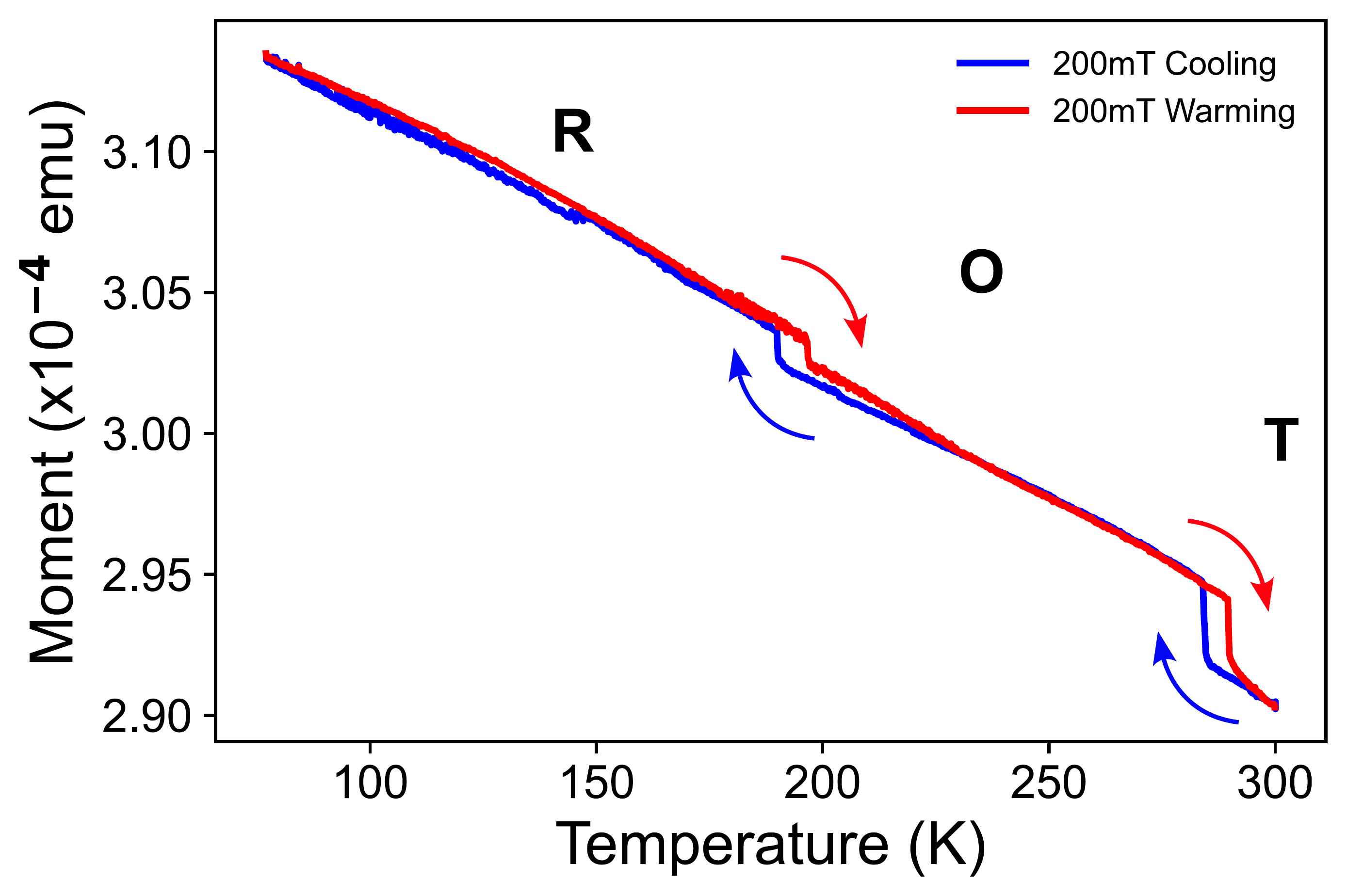}
    \caption{MvT measurement in an applied field of 200mT. The blue curve corresponds to data taken during cooling, with the red curve taken while heating to room temperature. The regions corresponding to the rhombohedral (\textbf{R}), orthorhombic (\textbf{O}), and tetragonal (\textbf{T}) phases of BaTiO$_3$ are indicated.}
    \label{MvT_200mT}
\end{figure}

We first study the change in magnetic moment with temperature using SQUID-VSM magnetometry. The temperature response of the sample is measured in an applied field of 200 mT between room temperature and 77 K. This temperature is well below the rhombohedral phase transition at 190 K. A magnetic field of this strength, greater than the local anisotropy field measured previously at room temperature, is chosen to bias the moment towards the magnetic field direction across phase transitions. The results in Fig. \ref{MvT_200mT} show two hysteretic changes in magnetic moment around 190 K and 280 K, corresponding to the rhombohedral-orthorhombic and orthorhombic-tetragonal phase transitions of the BTO respectively. This hysteretic behaviour is an expected property of the BTO substrates, which have been widely reported\cite{Acosta2017, Vaz2009, Venkataiah2012} to have hysteretic phase transitions in a variety of properties including the lattice parameters, which are closely coupled to the lattice elongations of the ferroelectric order. Similar behaviour has been observed in epitaxial BTO-ferromagnet systems in the literature and attributed to either the change in strain\cite{Vaz2009} or the change in ferroelectric domain structure (and so, the resulting magnetic easy axes) \cite{Venkataiah2012} and demonstrates strong strain coupling between the substrate and film in all crystal phases.

\begin{figure}
    \centering
    \includegraphics[width=\columnwidth]{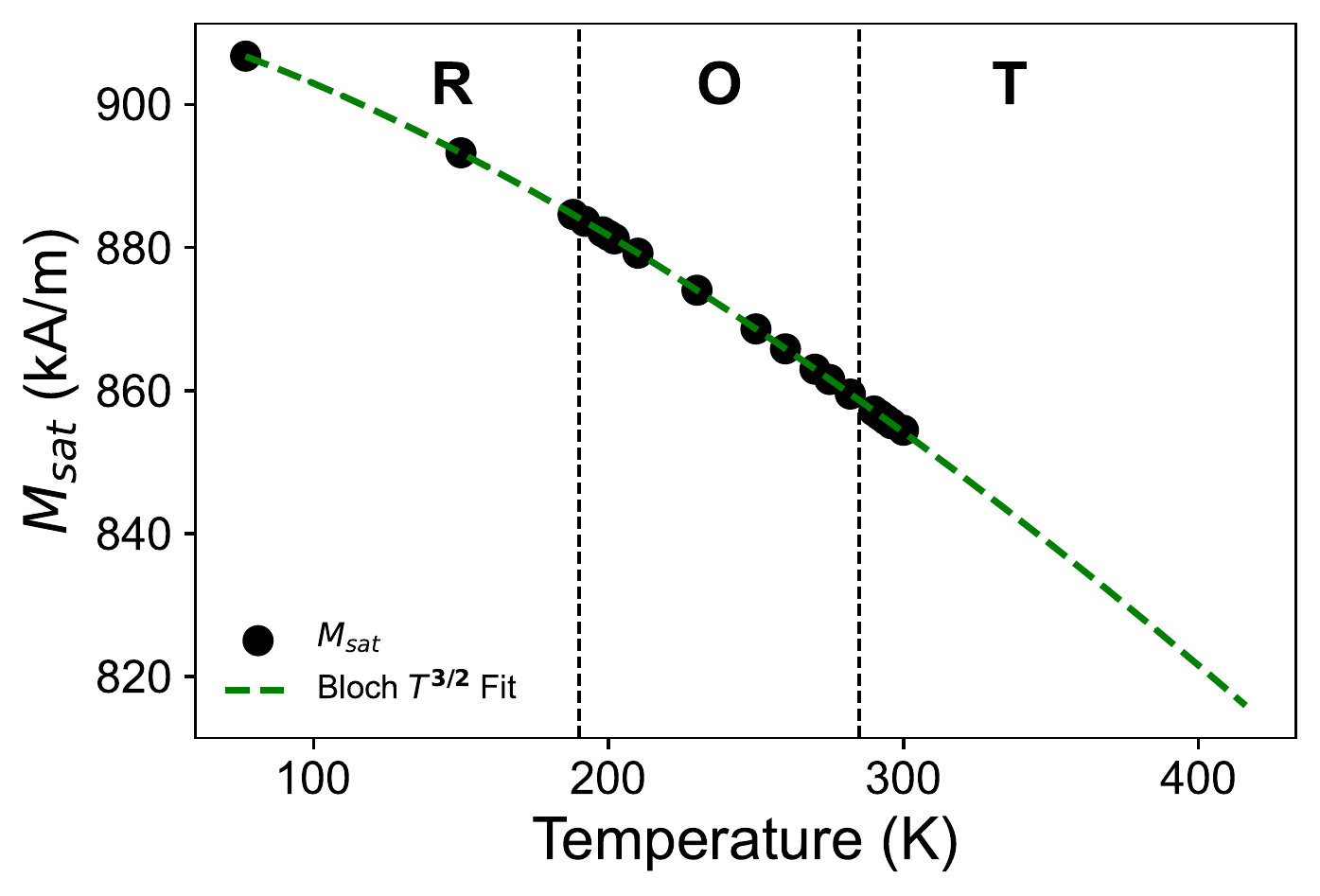}
    \caption{Saturation magnetization (points) for 20nm CoFeB thin film coupled to a BTO(111) substrate, extracted from hysteresis loops at each temperature and the corresponding Bloch $T^{3/2}$ law fit (dashed line). Indicated are the temperatures at which the rhombohedral (\textbf{R}), orthorhombic (\textbf{O}) and tetragonal (\textbf{T}) phase transitions occur. No significant change around these points is observed.}
    \label{Ms_T}
\end{figure}

Next we investigate the change in saturation magnetization with temperature. In Fig. \ref{Ms_T} we show the saturation magnetization as a function of temperature. The result of this is values that fit well to a standard $T^{3/2}$ fit with no discontinuities in $M_{s}(T)$ indicating that the strain is insufficient to change the saturation magnetization. This means that the temperature hysteresis observed in Fig. \ref{MvT_200mT} most likely corresponds to the abrupt change in ferroelectric domain structure at the phase transitions of the BTO substrate. While 200mT is in excess of the coercive field at all temperatures, it is not sufficient to pin the magnetization to the field direction as the ferroelectric domain structure, and consequently the local magnetoelastic anisotropy, abruptly changes. Performing the same MvT with a larger 1T field suppresses the change in moment across these transitions (see Supplementary Fig. 1). In the case of these amorphous thin-film ferromagnets, it would seem that the change in domain structure is more important than the change in strain for these whole-sample measurements.

\begin{figure*}
    \centering
    \includegraphics[width=2\columnwidth]{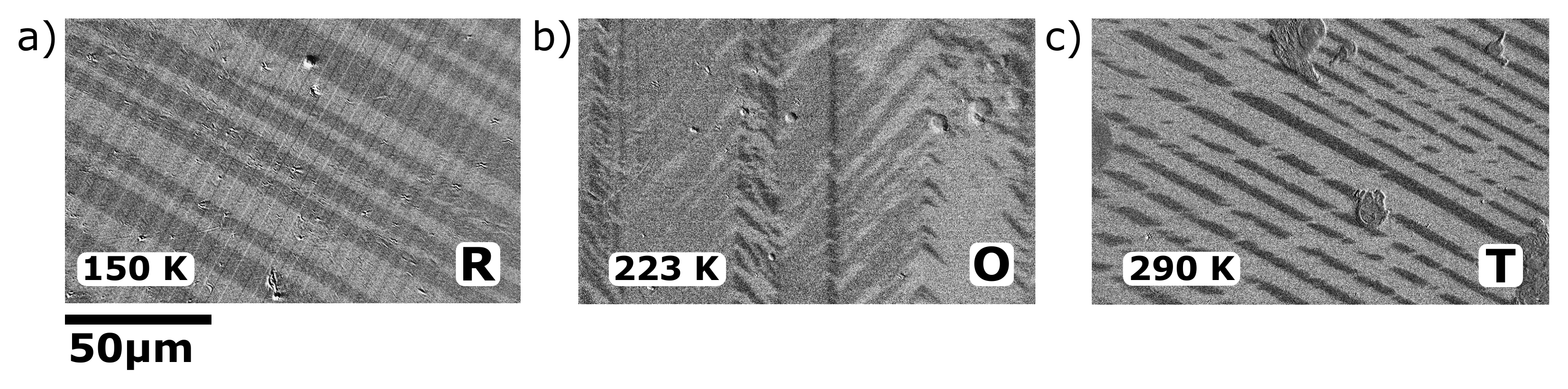}
    \caption{Ferromagnetic domain images taken in the same vicinity for the a) rhombohedral (\textbf{R}), b) orthorhombic (\textbf{O}), and c) tetragonal phases (\textbf{T}). The magnetic stripe orientation changes to match the ferroelectric domain structure demonstrating good coupling between films across phase transitions.}
    \label{ROT_imaging}
\end{figure*}

%% Discuss (111) elongations implications

The results from the SQUID-VSM measurements in Figs. \ref{MvT_200mT} and \ref{Ms_T} represent the volume-averaged response of the sample. As mentioned previously there are many ferroelectric domains within the sample with different orientations of lattice elongation that leads to a variety imprinted magnetoelastic easy axes in the CoFeB film. While we are able to infer strain coupling between the ferroelectric substrate and ferromagnetic film, it is not possible to probe in detail the effect of the substrate on the ferromagnetic film as we do not have a monodomain substrate, making it more challenging to interpret the results. To understand in more detail the effect of temperature on the strength of the magnetoelastic anisotropy within one ferroelectric domain we performed local measurements using a Kerr microscope to focus our measurements down to one region of ferroelectric domain structure and coupled regions corresponding to individual ferroelectric domains.

Using optical cryostat attachments we perform wide-field Kerr microscopy on the sample in two setups, a cryo-stage with liquid nitrogen that allows us to cool to 77 K (well below the orthorhombic and rhombohedral phase transitions) and a heater-stage in which we heat to above 420K (the tetragonal to cubic phase transitions). Fig. \ref{ROT_imaging} shows the result of domain imaging in the same region across three different phases, imaged here in zero applied magnetic field. The presence of stripe domains in all three polar phases are in excellent agreement with the SQUID data presented previously and work carried out on (100)-oriented substrates in the literature\cite{Lahtinen2013}, and shows that the strain is sufficient to couple the domains at all temperatures. 

In this experimental setup, it is not possible to precisely ascertain the ferroelectric (and anisotropic) configuration of the region before and after the phase transition as the sample cannot be freely rotated so a rigorous extrapolation of the easy axes cannot be performed. However, we can gain some understanding of the configuration from the orientation of the domain walls and the angle at which 180$^\circ$ domain walls are canted relative to the stripe axis. In the tetragonal and rhombohedral phases, the projection of lattice elongation onto the (111) surface lies in [11$\bar{2}$]-like directions and so we expect that stripes oriented in the same direction in both crystal phases will maintain the same ferroelectric configuration with the rotation between the easy axes remaining the same\cite{Hunt2023}. In the orthorhombic phase, the lattice elongation lies along [01$\bar{1}$]-like directions, which is the orientation of domain walls in the tetragonal and rhomobohedral phases. As a result, a rotation of the domain wall by 30$^\circ$ or 90$^\circ$ indicates that the ferroelectric configuration remains the same, and a rotation of 0$^\circ$ or 60$^\circ$ is indicative of a change in the ferroelectric configuration from either a 60$^\circ$ to 120$^\circ$ or vice versa. 

In the domain images presented here, the 180$^\circ$ reversal domains indicate that the \textbf{T}-phase region has a 60$^\circ$ configuration, and that the $\textbf{O}$-phase image has a 120$^\circ$ configuration, and indeed the stripe orientation rotates by 60$^\circ$ as we expect it should for a change in ferroelectric domain type. The stripe orientation is in the same crystal direction in the $\textbf{R}$ and $\textbf{T}$ phases from which we infer that the ferroelectric configuration is the same in both phases. We have shown the case where the ferroelectric configuration remains the same between the \textbf{R} and \textbf{T} phases - we have however also observed a rotation of 90$^\circ$ suggesting that both outcomes are equally likely. Separately, in the heater stage we observe that the magnetic stripe domain structure vanishes above the Curie temperature of BTO ($\approx$420K). 

Starting from a base temperature of 77 K the value of the anisotropy field is extracted from hard-axis hysteresis loops measured from individual stripe domains and the value of anisotropy calculated from $H_{k} = \frac{2K_{eff}}{\mu_{0}M_{s}}$. Values for $M_s$ are used from our previously presented data where possible and the obtained fit is used to extrapolate values for high temperatures.

\begin{figure}
    \centering
    \includegraphics[width=\columnwidth]{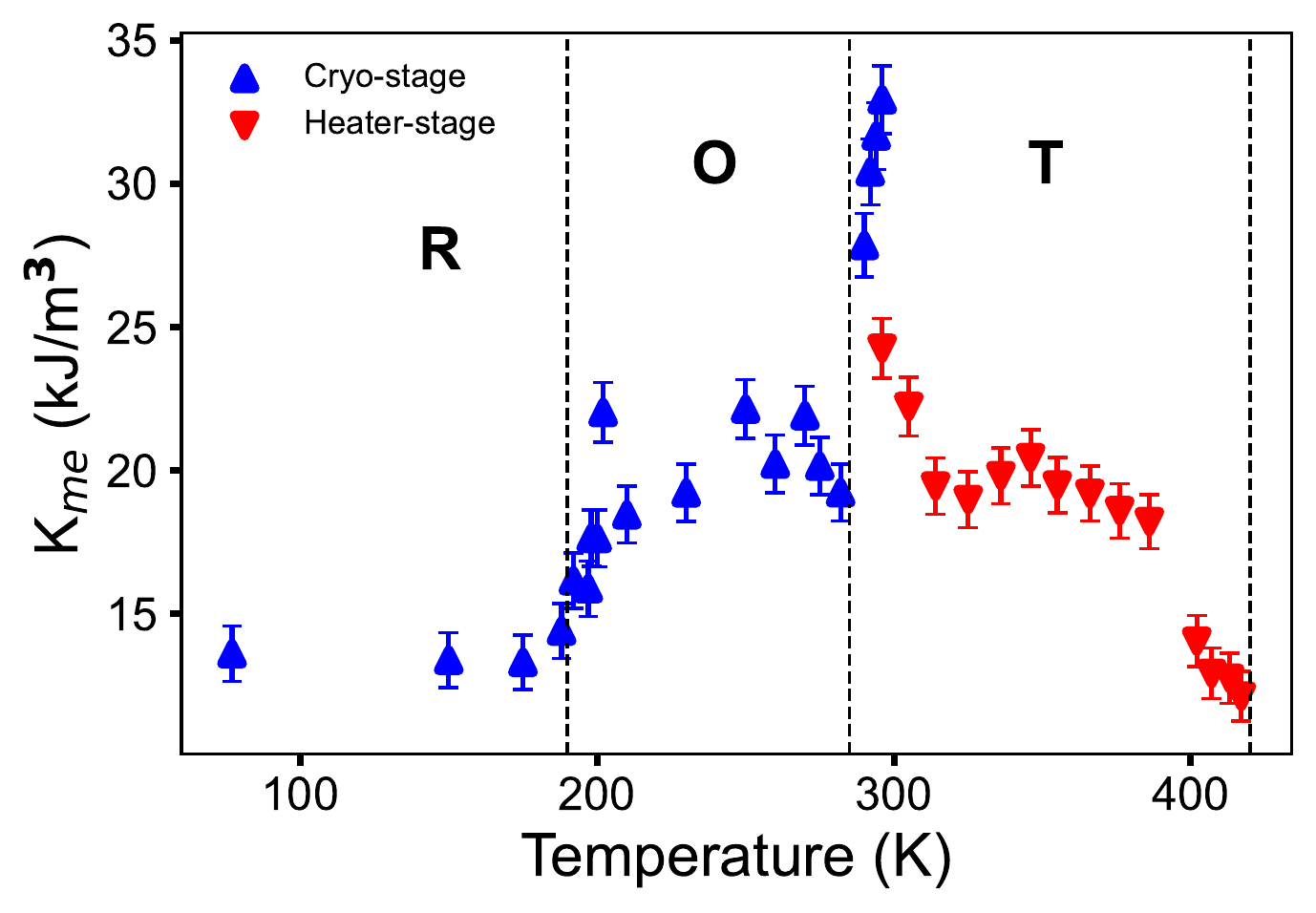}
    \caption{Local magnetoelastic anisotropy imprinted in the CoFeB film by the BaTiO$_3$(111) substrate. Dashed lines indicate the phase transitions between the rhombohedral (\textbf{R}), orthorhombic (\textbf{O}) and tetragonal (\textbf{T}) phases of the substrate. Measurements in blue (\textcolor{blue}{$\filledmedtriangleup$}) are taken at low temperature with the use of an optical cryostat, and measurements in red (\textcolor{red}{$\filledmedtriangledown$}) were taken separately in a heater stage.}
    \label{Kme}
\end{figure}

\begin{figure}
    \centering
    \includegraphics[width=\columnwidth]{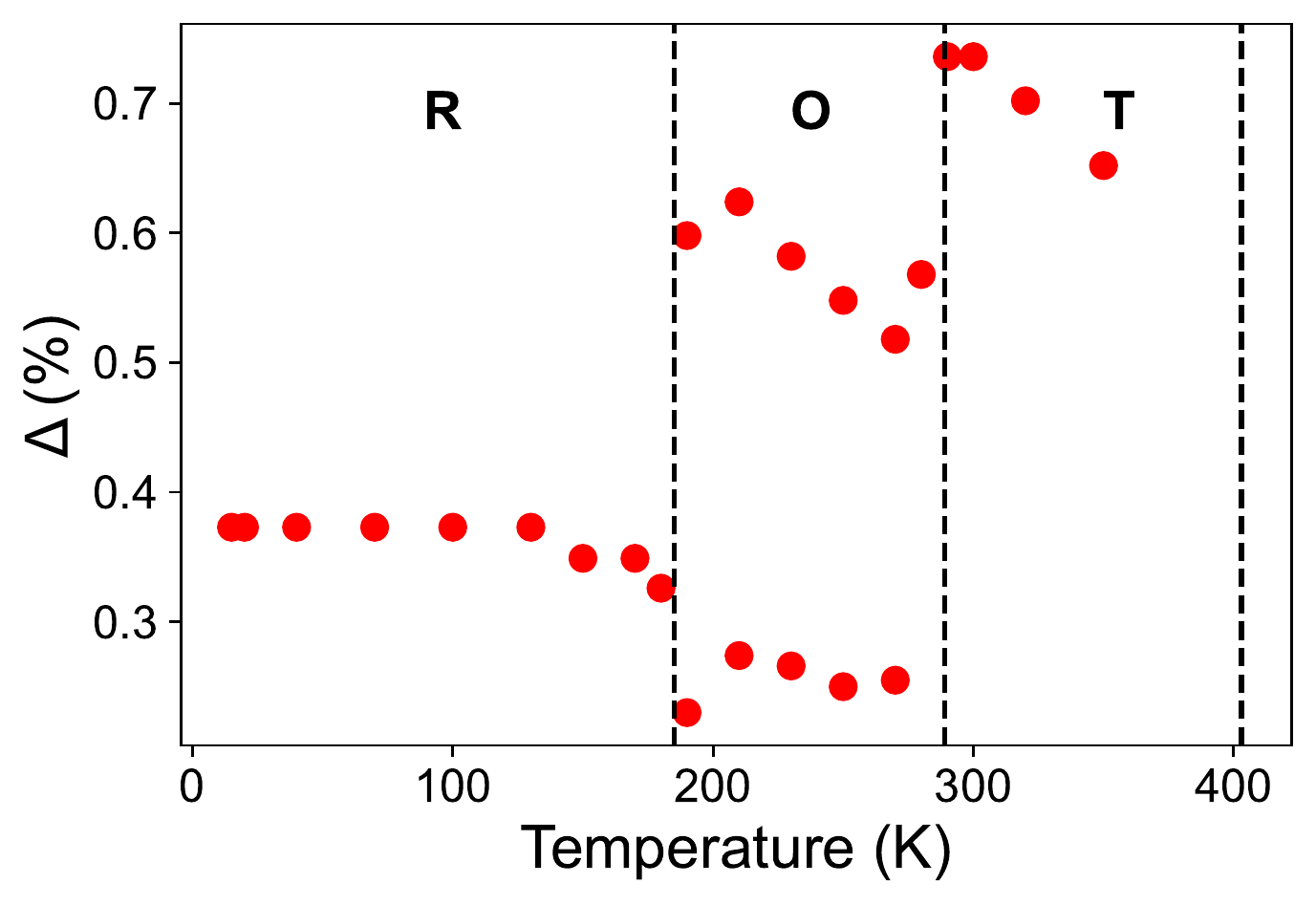}
    \caption{Calculated magnitude of the lattice elongations on the (111) plane using structural data obtained by Kwei et al\cite{Kwei1993}, expressed as a percentage of the in-plane lattice parameter.}
    \label{lattice-elong}
\end{figure}

The results are summarized in Fig. \ref{Kme}. The change in anisotropy far from the phase transitions shows a step-like change for the measurements performed in the cryo-stage, with different transition behaviour. From the rhombohedral to orthorhombic phase there is a change in the magnitude of the magnetoelastic anisotropy from an average value of (14$\pm$1) kJ/m$^3$ in the rhombohedral phase to a value of (20$\pm$1) kJ/m$^3$ in the orthorhombic phase. In the region around the O-R phase transition the magnitude of the magnetoelastic anisotropy smoothly transitions between these two average values. Whereas, when going from the orthorhombic phase to the tetragonal the increase is sharp with presumably some saturation to a maximum value around room temperature. To examine this in more detail we calculate in Fig. \ref{lattice-elong} the expected magnitude of the lattice elongations (relative to the in-plane lattice constant) in the (111) plane based on the reported results from Kwei et al\cite{Kwei1993}.

In the rhombohedral phase there is minimal change in the lattice parameters and indeed we observe no significant changes to the measured magnetoelastic anisotropy far from the orthorhombic phase transition. In the orthorhombic phase, we expect two distinct branches corresponding to directions of ferroelectric polarization with either the major or minor component of the biaxial lattice elongation lying in the (111) plane. We expect that the shorter lattice elongation should show a continuous change across the phase transition, however in our results the same behaviour is instead reflected in what must correspond to the larger lattice elongation as the magnetoelastic anisotropy increases from the rhombohedral to orthorhombic phase. The discontinuous `jump' in the lattice elongation at the orthorhombic-tetragonal transition is well reflected in our measurements of $K_{me}$.

For measurements performed on the heater stage, there is a large discrepancy that leads to a discontinuous transition between the two datasets. We attribute this to effects observed in previous work\cite{Hunt2023}, where there can be a large difference in the magnitude of K$_{me}$ between the 60$^\circ$ or 120$^\circ$ domain states, and measurements here were performed on different regions of the sample. Regardless, it shows the changes from the initial state that we expect with an approximately linear reduction from 315-385 K corresponding to the decrease in lattice elongations and then a sharp drop off close to the ferroelectric Curie temperature at approximately 420 K (in agreement with the Curie temperature seen in the work by Lahtinen et al\cite{Lahtinen2013}) where the polar order reduces. Above this Curie temperature, we do not observe any magnetic stripe domains as the ferroelectric order is no longer present and there is no meaningful in-plane anisotropy.

Using our measured values of $K_{me}$ we explore the resulting implications that the change in magnetoelastic anisotropy has for the magnetic domain wall width (DWW). In a micromagnetic\cite{Vansteenkiste2014} framework we have studied the effect of temperature on the DWW of charged and uncharged domain wall configurations. The simulation is divided into three distinct regions of uniaxial anisotropy corresponding to different directions of lattice elongation in adjacent ferroelectric domains. This is illustrated in Supplementary Fig. 2. The central stripe width is set to be 2 $\mu$m and the micromagnetic cell size is 2 nm x 2 nm x 20 nm, and periodic boundary conditions are used in the $x$ and $y$ directions. Values of $K_{u1}$, representing the parameter previously defined as $K_{me}$, and $M_{sat}$ are informed from the above experiment with the assumption made that similar magnitudes of $K_{me}$ can be expected for all ferroelectric domains. Values of $A_{ex}$ are taken from previous work examining the temperature dependence of exchange stiffness\cite{Moreno2016} of Co which is likely to be an overestimate for a CoFeB system. This is intended not to predict the exact values of the DWW but instead to examine how it scales with temperature taking into account the relative scaling of all relevant micromagnetic parameters. 

For all phases of BTO the projection of the polarization onto the (111) plane leads to an angle between adjacent lattice elongations of either 60$^\circ$ or 120$^\circ$, depending on which axes the polarization switches between. In each ferroelectric domain, the magnetization is strongly pinned to the imprinted magnetoelastic anisotropy axis which leads to magnetic domain walls that are pinned to the ferroelectric domain walls with a reduced wall angle dependent upon how the bulk magnetization rotates between the ferroelectric domains. The magnetization can rotate in a head-to-tail or head-to-head fashion leading to charged or uncharged domain wall structures. In total, this means that there are four domain wall structures to consider by combining the possible ferroelectric rotations and charged or uncharged states: a rotation of 60$^\circ$ with charged (60C) or uncharged (60U) character and a rotation of 120$^\circ$ with charged (C) or uncharged (U) character. These were previously illustrated in Fig. \ref{dw_configs}.

\begin{figure}
    \centering
    \includegraphics[width=\columnwidth]{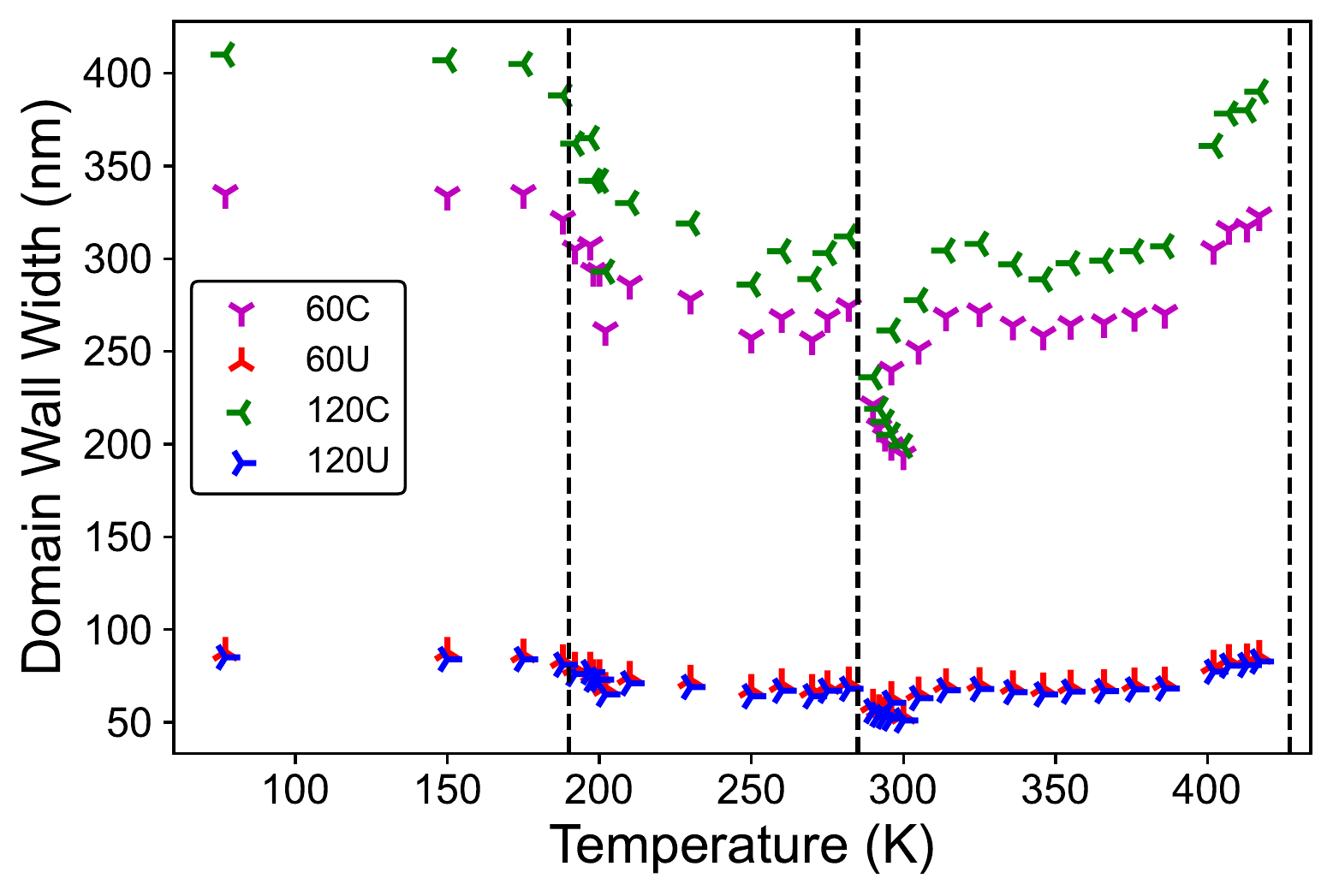}
    \caption{Domain wall width in the charged (C) and uncharged (U) configurations as a function of temperature. Numbers in the legend represent the total angle through the domain wall between the magnetoelastic anisotropy axes in adjacent stripe domains corresponding to the rotation of polarization.}
    \label{DWW}
\end{figure}

Within each of the pre-defined stripes, the magnitude of the in-plane anisotropy remains the same but the orientation of the magnetoelastic anisotropy is varied (rotating by either 60$^\circ$ or 120$^\circ$). Periodic boundary conditions are used in both the $x$ and $y$ dimensions and the domain wall is homogenous in the $y$-axis. To calculate the domain wall width, $\delta$, we use the method outlined in previous work\cite{Hunt2023, Franke2014} using the integral definition of the domain wall width:

\begin{equation}
    \delta = \int_{-\infty}^{\infty} \cos^{2}(\phi') \,dx ,
\end{equation}

with $\phi'$ being the reduced magnetization angle, 

\begin{equation}
    \phi' = \bigg(\phi-\frac{\mid\phi_{\frac{d}{2}} - \phi_{\frac{-d}{2}}\mid}
                {2}\bigg) 
            \frac{180}
                {\mid{\phi_{\frac{d}{2}} - \phi_{-\frac{d}{2}}\mid}},
\end{equation}

with $\phi$ being the magnetization angle measured relative to the direction of the easy axis and $\mid\phi_{\frac{d}{2}} - \phi_{\frac{-d}{2}}\mid$ being the change in magnetization angle between magnetic domains measured far from the domain wall.

The variation of each of these domain wall widths with temperature is shown in Fig. \ref{DWW}. In all cases, the scaling is most strongly dominated by the change in $K_{me}$ which has an inversely proportional change in the magnetic domain wall width. The result is a profile that shows the opposite temperature scaling to that shown in Fig. \ref{Kme}, with the very extremes of the measured temperatures (77 K, 417 K) displaying the largest values of domain wall width. The charged domain wall structures show the largest absolute change in DWW, representing something that would in principle be easier to measure using a high-resolution technique such as photo-emission electron microscopy to probe the domain wall profile. 

Differences resulting from the different magnetoelastic configurations become apparent only at the temperature extremes with changes in all uncharged configurations being extremely small, on the order of a few nanometers of difference, while the difference between the 60C and 120C is on the order of 50nm at the largest. The differences at these extremes stem only from the change in angle between the easy axes, indicating that the charged domain wall width could be a useful way of measuring the two distinct states. 

\section{Conclusion}

In summary, we have studied the temperature dependence of BaTiO$_3$(111)/CoFeB heterostructures using SQUID-VSM and Kerr microscopy and determined the change in magnetoelastic anisotropy as a result of temperature-dependent strain transfer. We find strong coupling between the substrate and thin film at all temperatures, with the local anisotropy varying as the lattice parameters of the BaTiO$_3$ change with temperature. The absolute changes in the magnetoelastic anisotropy either side of room temperature are approximately equal, and so this presents two routes by which its magnitude could be changed in a device. Following this, we performed simulations informed by our experiments to determine how this dependence dominates the domain wall width of magnetic domains. These domain wall widths were found to be most strongly dependent on the magnetoelastic anisotropy, with the low-anisotropy charged domain walls showing a large difference in domain wall widths between the 60$^\circ$ and 120$^\circ$ - a result purely of the difference in anisotropy configuration and resulting wall angles of the magnetic domain wall structure. 

These measurements demonstrate two distinct regions of interest around the rhombohedral-orthorhombic and orthorhombic-tetragonal transitions with the low-temperature results showing step-wise changes in the magnitude of the anisotropy, and high-temperature results demonstrating a drop off as the polar ordering weakens. Both routes to manipulating the magnetic anisotropy lead to similar absolute changes in the magnitude of the anisotropy. Control of magnetoelastic anisotropy is fundamental to devices based on multiferroic heterostructures. The anisotropy tuning that we show here will be useful for future devices based on BaTiO$_3$(111).

\begin{acknowledgments}

We acknowledge support from the Henry Royce Institute and funding from the Engineering and Physical Sciences Research Council (EPSRC) Grant No. EP/M000923/1. Variable-temperature, wide-field Kerr microscopy was performed at the Exeter Time-Resolved Magnetism Facility (EXTREMAG) (EPSRC Grant Reference EP/R008809/1 and EP/V054112/1). This project received funding in part from the European Union’s Horizon 2020 research and innovation programme under the Marie Sklodowska-Curie grant agreement No 750147. R. G. H. acknowledges the support of an EPSRC DTA studentship. 

\end{acknowledgments}

\section*{Author Declarations}

\section*{Conflict of interest}
    The authors have no conflicts to disclose.

\section*{Author Contributions}

\textbf{R. G. Hunt}: Investigation (lead); methodology (lead); conceptualization (equal); writing - original draft (lead); visualization (lead); formal analysis (lead); writing - review and editing (equal). \textbf{K. J. A. Franke}: Conceptualization (equal); formal analysis (supporting); writing - review and editing (equal). \textbf{P. S. Keatley}: Investigation (supporting), Methodology (supporting), writing - review and editing (equal). \textbf{P. M. Shepley}: Investigation (supporting); methodology (supporting), writing - review and editing (equal). \textbf{M. Rogers}: Investigation (supporting); methodology (supporting); writing - review and editing (equal). \textbf{T. A. Moore}: Project administration (lead); funding acquisition (lead); conceptualization (equal); methodology (supporting); investigation (supporting); writing - review and editing (equal).

\section*{Data Availability Statement}

The data associated with this paper is available from the Leeds research depository at https://doi.org/10.5518/1349.

\bibliography{BTO111-CoFeB}% Produces the bibliography via BibTeX.

\begin{thebibliography}{25}
\bibitem [{Ext()}]{Extremag}%
  \BibitemOpen
  \href@noop {} {}\bibinfo {note} {The Exeter time-resolved magnetism
  (EXTREMAG) facility is an EPSRC funded ultrafast laser facility that offers
  its users low temperature and high magnetic field environments in a variety
  of time-domain optical pump-probe, microscopy, and terahertz
  experiments.}\BibitemShut {Stop}%
\end{thebibliography}

\end{document}